%% file: main.tex
\documentclass[sigconf]{acmart}

\usepackage{bigstrut}
\usepackage{booktabs}
\usepackage{multirow}
\usepackage{microtype}
\usepackage{graphicx}
\usepackage{subfig}
\usepackage{overpic}
\usepackage{balance}
\usepackage{amsthm}
\usepackage{array}
\usepackage{clrscode}
\usepackage{multicol}
\usepackage{float}
\usepackage{newfloat}
\usepackage{color}
\usepackage{xcolor}
\usepackage{amsopn}
\usepackage{mathrsfs}
\usepackage{mathtools}
\usepackage{amsmath}
\usepackage{arydshln}
\usepackage{blkarray}
\usepackage{enumerate}
\usepackage{courier}
\usepackage{rotating}
\usepackage{bm}
\usepackage{wrapfig}
\usepackage{ragged2e}
\usepackage[misc]{ifsym}
\usepackage[most]{tcolorbox}
\usepackage{threeparttable}
\usepackage{makecell}
\usepackage{adjustbox} 
\usepackage{enumitem}
\usepackage{hyperref}
\usepackage{url}
\usepackage{xspace}
\usepackage{comment}
\usepackage{changepage}
\usepackage{algorithm}
\usepackage{algpseudocode} 
\usepackage{algorithmicx} 
\usepackage[ruled, vlined, nofillcomment, linesnumbered, algo2e]{algorithm2e}

\definecolor{codegreen}{rgb}{0,0.3,0.6}
\definecolor{codegray}{rgb}{0.5,0.5,0.5}

\DeclareMathOperator*{\argmin}{arg\,min}

\newcommand{\ie}{\emph{i.e.,}\xspace}
\newcommand{\eg}{\emph{e.g.,}\xspace}

\newcommand{\paratitle}[1]{\vspace{1.5ex}\noindent\textbf{#1}}
\newcommand{\wrt}{w.r.t.\xspace}

\newcommand{\ignore}[1]{}

\AtBeginDocument{%
  \providecommand\BibTeX{{%
    \normalfont B\kern-0.5em{\scshape i\kern-0.25em b}\kern-0.8em\TeX}}}

\setcopyright{acmlicensed}
\copyrightyear{2018}
\acmYear{2018}
\acmDOI{XXXXXXX.XXXXXXX}

\acmConference[Conference acronym 'XX]{Make sure to enter the correct
  conference title from your rights confirmation emai}{June 03--05,
  2018}{Woodstock, NY}
\acmISBN{978-1-4503-XXXX-X/18/06}

\begin{document}

\title{Universal Item Tokenization for Transferable Generative Recommendation}

\author{Bowen Zheng}
\orcid{0009-0002-3010-7899}
\affiliation{%
    \institution{
    Gaoling School of Artificial Intelligence,  
    Renmin University of China}
    \city{Beijing}
    \country{China}
}
\email{bwzheng0324@ruc.edu.cn}

\author{Hongyu Lu}
\affiliation{%
    \institution{WeChat, Tencent}
    \city{Guangzhou}
    \country{China}
}
\email{luhy94@gmail.com}

\author{Yu Chen}
\affiliation{%
    \institution{WeChat, Tencent}
    \city{Beijing}
    \country{China}
}
\email{nealcui@tencent.com}

\author{Wayne Xin Zhao
\textsuperscript{\Letter}
}
\orcid{0000-0002-8333-6196}
\affiliation{
    \institution{
    Gaoling School of Artificial Intelligence,  
    Renmin University of China}
    \city{Beijing}
    \country{China}
}
\email{batmanfly@gmail.com}

\author{Ji-Rong Wen}
\orcid{0000-0002-9777-9676}
\affiliation{
    \institution{
    Gaoling School of Artificial Intelligence,  
    Renmin University of China}
    \city{Beijing}
    \country{China}
}
\email{jrwen@ruc.edu.cn}

\thanks{\Letter \ Corresponding author.}

\renewcommand{\shortauthors}{Bowen Zheng, et al.}

\begin{abstract}

Recently, generative recommendation has emerged as a promising paradigm, attracting significant research attention.
The basic framework involves an item tokenizer, which represents each item as a sequence of codes serving as its identifier, and a generative recommender that predicts the next item by autoregressively generating the target item identifier.
However, in existing methods, both the tokenizer and the recommender are typically domain-specific, limiting their ability for effective transfer or adaptation to new domains.
To this end, we propose \textbf{UTGRec}, a \underline{U}niversal item  \underline{T}okenization approach for transferable \underline{G}enerative  \underline{Rec}ommendation.
Specifically, we design a universal item tokenizer for encoding rich item semantics by adapting a multimodal large language model (MLLM). 
By devising tree-structured codebooks, we discretize content representations into corresponding codes for item tokenization.
To effectively learn the universal item tokenizer on multiple domains, we introduce two key techniques in our approach.
For raw content reconstruction, we employ dual lightweight decoders to reconstruct item text and images from discrete representations to capture general knowledge embedded in the content.
For collaborative knowledge integration, we assume that co-occurring items are similar and integrate collaborative signals through co-occurrence alignment and reconstruction.
Finally,  we present a joint learning framework to pre-train and adapt the transferable generative recommender across multiple domains.
Extensive experiments on four public datasets demonstrate the superiority of UTGRec compared to both traditional and generative recommendation baselines.
Our code is available at \textcolor{blue}{\url{https://github.com/RUCAIBox/UTGRec}}.

\end{abstract}



\maketitle

\input{sec1_introduction}

\input{sec2_methodology}

\input{sec3_experiment}

\input{sec4_relatedwork}

\input{sec5_conclusion}


\bibliographystyle{ACM-Reference-Format}
\balance

\bibliography{ref}

\end{document}

%% file: sec1_introduction.tex
\section{Introduction}
\label{sec:introduction}

Sequential recommender systems aim to capture a user's personalized preference based on his/her historical interaction records.
Typically, traditional approaches~\cite{fpmc,fossil} assign a unique ID to each item and represent user historical interactions as an item ID sequence in chronological order.
Various methods~\cite{bert4rec,gru4rec,sasrec,s3rec,fmlp-rec} have been proposed to model item ID sequences and predict the next item the user is likely to interact with. 

Recently, motivated by the promising potential of the generative paradigm in large language models (LLMs)~\cite{llm_survey} and generative retrieval methods~\cite{dsi,nsi,genret,autoindexer}, a number of studies have explored this paradigm within recommender systems~\cite{gptrec,tiger,howtoindex,lc-rec,mbgen,actionpiece}.
The basic idea of generative recommendation is to represent an item using a sequence of codes as its identifier instead of a single vanilla ID. 
Then, a generative model (\eg T5~\cite{t5}) is employed to autoregressively generate the target item identifier, thereby achieving next-item prediction in a sequence-to-sequence manner. 
In this paradigm, the process of mapping an item to its identifier, referred to as \emph{item tokenization}, plays a crucial role.
Existing methods for item tokenization include co-occurrence matrix decomposition~\cite{gptrec}, hierarchical clustering~\cite{seater,howtoindex}, and multi-level vector quantization~\cite{tiger,letter,lc-rec}.
The learned tokenizer associates each item with a series of codes that imply semantic knowledge, with shared prefix codes among different items reflecting their semantic similarity.
To further enhance the item tokenizer, some studies~\cite{letter,etegrec} propose to incorporate collaborative signals to improve the quality of item identifiers and their effectiveness in recommendation tasks.

Despite the effectiveness, existing item tokenization methods are typically developed in a \emph{domain-specific} manner. 
They often perform item clustering or train a RQ-VAE~\cite{rqvae} using pre-encoded item embeddings derived from a single domain~\cite{tiger,seater}, making the learned tokenizer the generative recommendation model less transferable across domains. 
Although it seems intuitive to directly extend these methods by training with multi-domain data, simple algorithms (\eg K-means) or model architectures (\eg RQ-VAE with MLP encoder) struggle to effectively capture the diverse and complex semantics across multiple domains~\cite{recformer,missrec}.  
Moreover, pre-encoded item embeddings cannot be jointly trained with the tokenizer model, limiting the model's ability to fully exploit its potential in learning item semantics. 
Considering the above limitations, we aim to develop a universal item tokenizer that can well transfer across multiple domains and also adapt to new domains.
To achieve this, two key technical challenges should be addressed. First, the extensive variety of items across multiple domains entails diverse and complex semantics, necessitating the development of a tokenizer model with robust capabilities in content understanding and generalization.
Second, it is important to effectively learn the item semantics while simultaneously integrating collaborative knowledge in recommendation scenarios. 

In this paper, we propose \textbf{UTGRec}, a \underline{U}niversal item  \underline{T}okenization approach for transferable \underline{G}enerative  \underline{Rec}ommendation.
Different from existing domain-specific item tokenization methods~\cite{tiger,letter}, our approach leverages the multimodal content of items as input and pre-trains a tokenizer model across multiple domains. 
Specifically, we focus on two key aspects, namely developing a universal item tokenizer and pre-training the model through item content reconstruction with collaborative integration.
For universal item tokenization, we first learn multimodal token representations for encoding essential item semantics based on a multimodal large language model (MLLM). 
Then, we propose tree-structured codebooks for representation discretization, which enhances multi-domain semantic fusion through codebook sharing.
For tokenizer pre-training,  we propose an optimization objective that introduces two key techniques, namely raw content reconstruction and collaborative knowledge integration, for jointly capturing item and collaborative semantics. 
Raw content reconstruction introduces dual lightweight decoders to reconstruct the raw text and image from discrete representations, thereby learning the multi-modal knowledge embedded in the content.
Regarding collaborative knowledge integration, we follow the assumption that co-occurring items are similar and incorporate collaborative knowledge into the item tokenizer via co-occurring item alignment and reconstruction.
A key merit of our method lies in that it can jointly learn multimodal representation encoding and discretization modules.
Finally, given the universal item tokenizer, we propose a learning framework to pre-train a transferable generative recommender.

In summary, our key contributions are as follows:

$\bullet$ We propose UTGRec, a transferable generative recommendation framework that can leverage both multimodal item content and collaborative knowledge for universal item tokenization. 

$\bullet$  We design a novel representation discretization method via tree-structured codebooks while presenting an optimization approach that integrates content reconstruction with co-occurring item alignment and reconstruction.

$\bullet$ We conduct extensive experiments on four public datasets to evaluate the effectiveness of our approach, demonstrating that UTGRec outperforms all baselines and attains significant improvements in transferable recommendation.

%% file: sec2_methodology.tex
\begin{figure*}[]
\centering
\includegraphics[width=0.98\linewidth]{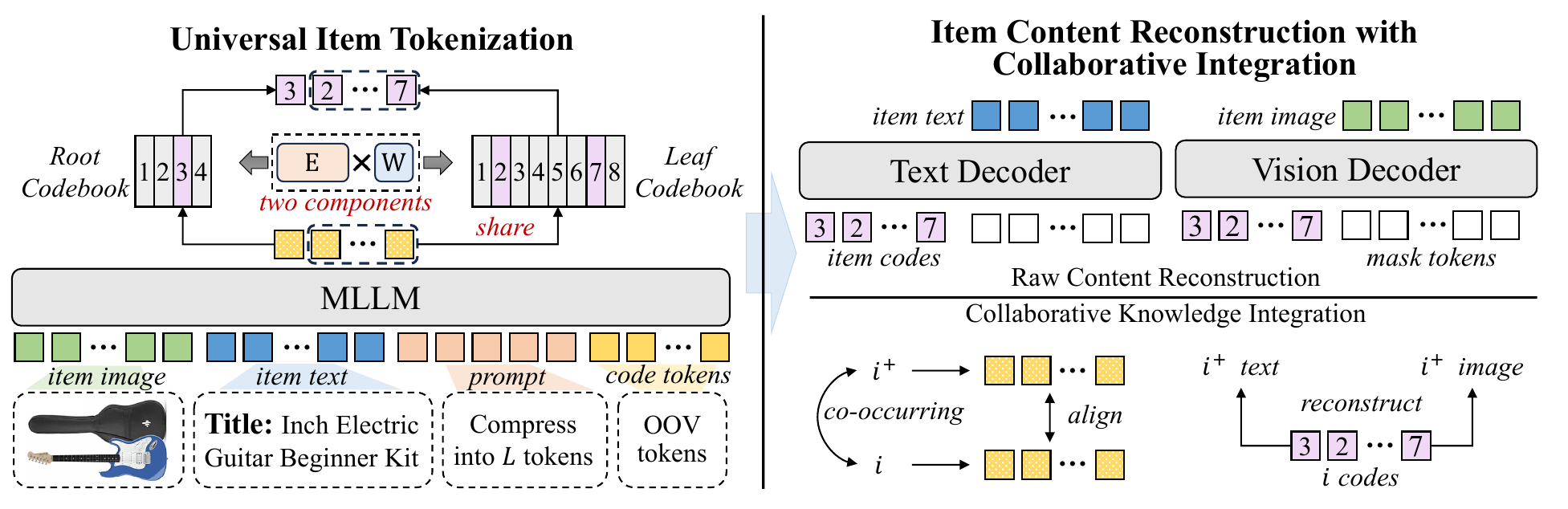}
\caption{The overall framework of UTGRec.}
\label{fig:model}
\end{figure*}

\section{Methodology}
\label{sec:methodology}
In this section, we present a universal item tokenization approach for transferable generative recommendation, named \textbf{UTGRec}.

\subsection{Problem Formulation and Overview}
\label{sec:problem}

Following prior studies~\cite{unisrec,missrec}, we consider the sequential recommendation task across multiple domains.
For each domain, an item set $\mathcal{I}$ is given, and a user's historical interactions are denoted as an item ID sequence $S = [i_1, i_2, \dots, i_n]$ arranged in chronological order, where $i \in \mathcal{I}$ denotes an interacted item.
Sequential recommendation aims to capture user preferences and predict the next potential item $i_{n+1}$ based on $S$.
Different from the conventional setting, we adopt the generative paradigm in which each item is associated with a list of codes $[c_1,\dots,c_L]$ derived from its text $T_i$ and image $V_i$, serving as its identifier.
$L$ is the length of the identifier.
Through the above process called item tokenization, the item sequence $S$ and the target item $i_{n+1}$ can be tokenized into a pair of code sequences $X = [c^1_1,c^1_2, \dots, c^n_{L-1},c^n_L]$ and $Y = [c^{n+1}_1,\dots,c^{n+1}_L]$, where each item is represented by $L$ codes.
Thus, the traditional sequential recommendation task is reformulated as a sequence-to-sequence problem, in which the next-item prediction is achieved by autoregressively generating the target item identifier (\ie $Y$).
Formally, this problem can be written as:
\begin{align}
    P(Y|X) = \prod_{l=1}^L  P(c^{n+1}_l|X,c^{n+1}_1,\dots,c^{n+1}_{l-1}).
\end{align}

To assign item identifiers, previous studies~\cite{tiger,gptrec,letter} have developed various item tokenizers. 
However, these tokenizers are typically domain-specific, generative recommenders trained with such tokenizers lack the ability to generalize to new domains.
In this paper, we address this limitation by learning a universal item tokenizer.
The overall framework is illustrated in Figure~\ref{fig:model}. 
First, we develop a universal item tokenizer based on the multimodal content of items and propose tree-structured codebooks for representation discretization (Section~\ref{sec:uni_token}).
Second, we perform item content reconstruction via discrete representations for multimodal semantic learning and integrate collaborative knowledge through alignment and reconstruction of co-occurring items (Section~\ref{sec:reconstruction}). 
Finally, we present a transferable generative recommender learning framework based on the universal item tokenizer (Section~\ref{sec:training}).

\subsection{Universal Item Tokenization}
\label{sec:uni_token}

To develop the universal item tokenizer, a key step is to sufficiently leverage the rich item semantics and establish a representation scheme that can well transfer across domains. Unlike previous text-based transferable item representations~\cite{unisrec,vqrec}, we consider leveraging multimodal item content for capturing more comprehensive semantic information. 
Moreover, to achieve multi-domain item tokenization, we design a novel representation discretization approach via tree-structured codebooks.

\ignore{In order to achieve transferability, a common approach is to leverage the content information of items, such as natural language text and visual images, to extract universal knowledge across different domains~\cite{unisrec,missrec}. 
While some prior studies~\cite{tiger,letter} utilize pre-encoded item text representations for item tokenization, they typically train item tokenizers on items from a single domain.
Furthermore, non-optimizable item representations and weak model architectures (\eg RQ-VAE with MLP encoder) limit the ability of these methods to model the diverse and complex semantics present across multiple domains.
In contrast, we apply the multimodal information of items as the semantic bridge and adopt an optimizable MLLM as the backbone to achieve universal item tokenization.
}

\subsubsection{Item Content Encoding via MLLM}
\label{sec:encode}

In our framework, each item $i$ is associated with textual information $T_i$ (\eg the item title, features, and category) and a corresponding image $V_i$. 
To efficiently encode the multimodal content, we employ a MLLM to compress this information into several representations.
Specifically, we employ Qwen2-VL~\cite{qwen2-vl} as the backbone model and formalize the item's textual and visual information as the following input prompt:

\begin{center}
\begin{tcolorbox}[
colback=black!5!white,
colframe=codegray,
width=0.45\textwidth,
title={Prompt for Item Content Encoding}
]
{
<|vision\_start|><IMAGE><|vision\_end|> \\
This is an image of an item. The item's information also includes: ``<TEXT>''. Compress the image and text information into $L$ tokens, representing content from coarse to fine granularity. \\
<CODE\_$1$><CODE\_$2$> $\cdots$ <CODE\_$L$>
}
\end{tcolorbox}
\end{center}

In this prompt, ``<|vision\_start|>'' and ``<|vision\_end|>'' are special tokens that delineate the beginning and end of the vision input.
``<IMAGE>'' is a placeholder for the image and would be replaced by visual token embeddings derived from the raw image $V_i$. 
Similarly, ``<TEXT>'' is a placeholder for the text $T_i$.
Furthermore, ``<CODE\_$1$> $\cdots$ <CODE\_$L$>'' represent $L$ code tokens used to integrate content information, which are added to the MLLM vocabulary.
Formally, we denote this prompt as $P(V_i,T_i)$ and feed it into the MLLM to obtain item representations:
\begin{align}
    \mathbf{H} = [\bm{h}_1, \bm{h}_2,\dots, \bm{h}_L] = \text{MLLM}(P(V_i,T_i)),
    \label{eq:mllm_enc}
\end{align}
where $\mathbf{H} \in \mathbb{R}^{L \times d}$ denotes the output hidden states corresponding to tokens code tokens, serving as the representations of the compressed item content information.

\subsubsection{Representation Discretization via Tree-Structured Codebooks}
\label{sec:tree_code}
After obtaining $L$ content representations of an item, we need to map them into discrete codes for item tokenization. 
A straightforward method is to learn multi-level codebooks, as proposed in existing studies~\cite{tiger,letter}, to discretize each content representation individually.
However, this method presents two potential issues.
First, although we instruct the MLLM to compress information of different granularities into multiple representations, these representations are essentially derived from the same item content, resulting in high similarity.
Consequently, items with the same prefix code often share subsequent codes.
Secondly, codebook representation collapse~\cite{DBLP:journals/corr/abs-1805-11063} is more likely to occur across multiple domains, that is, only a few codebook vectors are activated in each domain.
This phenomenon significantly affects multi-domain fusion and undermines the sufficient utilization of codebooks.
To address the above issues, we first apply the prefix residual operation to transform content representations into a basic representation and multi-level incremental representations, thereby alleviating high similarity, 
Subsequently, we present tree-structured codebooks, which enable joint optimization of codebook space and enhance codebook utilization by sharing multi-level codebooks and parameterizing each codebook into two components.

\paratitle{Prefix Residual Operation.}
Given the content representations $\mathbf{H}$, we leverage a MLP to project it to $\mathbf{H}^{\prime} \in \mathbb{R}^{L \times d_c}$ for dimension adaptation, where $d_c$ denotes the codebook dimension. 
Afterward, we apply the prefix residual operation as follows: 
\begin{align}
    \tilde{\bm{h}}_l = \bm{h}^{\prime}_l - \bm{h}^{\prime}_{l-1}, \quad l \in {2, \dots, L},
\end{align} 
where $\bm{h}^{\prime}_l$ denotes the $l$-th representation in $\mathbf{H}^{\prime}$.
Unlike the recursive subtraction used in residual quantization~\cite{rq}, our approach extracts the incremental representation $\tilde{\bm{h}}_l$ that captures the incremental information relative to the prefix representation, which is more diverse and distinguishable.
Furthermore, $\tilde{\bm{h}}_1= \bm{h}^{\prime}_1$ serves as the item's basic content representation, which remains unchanged.

\paratitle{Tree-Structured Codebooks.}
With the above prefix residual operation, we can obtain the basic representation and multi-level incremental representations of an item. 
To discretize these representations into a list of codes as the item identifier, we propose tree-structured codebooks, which consist of two components: the root codebook and the leaf codebook. 
The root codebook is dedicated to the basic representation, capturing the primary content information of the item. 
The leaf codebook is shared across the multi-level incremental representations to model the incremental information from multiple domains, which can improve the codebook utilization and enhance multi-field fusion~\cite{DBLP:conf/cvpr/LeeKKCH22}.

To alleviate representation collapse, we parameterize each codebook as $\mathbf{E}\mathbf{W}$, taking inspiration from recent work~\cite{simvq}. Here, $\mathbf{E} \in \mathbb{R}^{K \times d_c}$ denotes the codebook matrix and $\mathbf{W} \in \mathbb{R}^{d_c \times d_c}$ is a projection matrix, where $K$ denotes the size of the codebook.
Unlike traditional codebooks that only involve the codebook matrix $\mathbf{E}$, this method optimizes each row of the codebook matrix $\mathbf{E}$ while jointly updating the projection matrix $\mathbf{W}$, thereby driving the optimization of the entire codebook space~\cite{simvq}.
This advantage allows the gradient from each domain to drive the update of the entire codebook space, effectively preventing the fragmentation of codebook vectors and promoting multi-domain fusion.
Then, we denote the root and leaf codebooks as $\mathbf{C}^r = \mathbf{E}^r\mathbf{W}^r$ and $\mathbf{C}^f = \mathbf{E}^f\mathbf{W}^f$ respectively and employ them to discretize representations as follows:
\begin{align}
c_l = 
\begin{cases}
    \underset {j} { \argmin } ||\tilde{\bm{h}}_l - \bm{e}_j^r \mathbf{W}^r||_2^2, & \text{if } l=1, \\
    \underset {j} { \argmin } ||\tilde{\bm{h}}_l - \bm{e}_j^f\mathbf{W}^f||_2^2, & \text{if } l > 1,
\end{cases}
\end{align}
where $c_l$ denotes the $l$-th code of the item, $\bm{e}_j^r$ and $\bm{e}_j^f$ are the $j$-th codebook vectors in $\mathbf{E}^r$ and $\mathbf{E}^f$ respectively.
The optimization objective of tree-structured codebooks is formulated as follows:
\begin{align}
    & \mathcal{L}_{\text{Code}} = \frac{1}{2} (\mathcal{L}^r + \mathcal{L}^f), 
\end{align}
where $\mathcal{L}^r$ and  $\mathcal{L}^f$ are defined as follows: 
\begin{align}
    & \mathcal{L}^r = ||\operatorname{sg}[\tilde{\bm{h}}_1] - \bm{e}_{c_1}^r\mathbf{W}^r||_2^2 + \beta \ ||\tilde{\bm{h}}_1 - \operatorname{sg}[\bm{e}_{c_1}^r\mathbf{W}^r]||_2^2, \\
    & \mathcal{L}^f = \frac{1}{L-1}\sum_{l=2}^{L} ||\operatorname{sg}[\tilde{\bm{h}}_l] - \bm{e}_{c_l}^f\mathbf{W}^f||_2^2 + \beta \ ||\tilde{\bm{h}}_l - \operatorname{sg}[\bm{e}_{c_l}^f\mathbf{W}^f]||_2^2,
    \label{eq:code_loss}
\end{align}
where $\operatorname{sg}[\cdot]$ denotes the stop-gradient operation.
$\beta$ is used to balance the optimization between the item representations and codebooks, typically set to 0.25.
$\mathcal{L}^r$ and $\mathcal{L}^f$ represent the losses associated with the root and leaf codebooks, respectively.


\subsection{Item Content Reconstruction with Collaborative Integration}
\label{sec:reconstruction}
Unlike existing works~\cite{tiger,letter} that learn codebook semantics based on pre-coded embeddings, we employ the raw content information as reconstruction targets to ensure the completeness of the information.
Simultaneously, we jointly optimize the item encoding and representation discretization process to achieve better multi-domain content understanding.
Moreover, since simple content reconstruction cannot integrate collaborative knowledge in recommender systems, we further align and reconstruct co-occurring items to achieve collaborative integration within item tokenizer.

\subsubsection{Raw Content Reconstruction}
Based on tree-structured codebooks, we can obtain the codes of the item $[c_1,\dots,c_L]$ and corresponding discrete representations $[\bm{e}_{c_1}^r\mathbf{W}^r, \bm{e}_{c_2}^f\mathbf{W}^f, \dots,\bm{e}_{c_L}^f\mathbf{W}^f]$. 
Then, we apply the inverse of the prefix residual operation, recursively restoring discrete increment representations into discrete content representations.
Formally, the process can be written as: $ \hat{\bm{h}}_l = \hat{\bm{h}}_{l-1} + \bm{e}_{c_l}^f\mathbf{W}^f$, where $\hat{\bm{h}}_1 = \bm{e}_{c_1}^r\mathbf{W}^r $, and the $l$-th discrete content representation $\hat{\bm{h}}_l$ is obtained by adding the $l$-th discrete increment representation $\bm{e}_{c_l}^f\mathbf{W}^f$ to the last discrete content representation $\hat{\bm{h}}_{l-1}$.
After a linear projection for dimension adaptation, we denote all discrete content representations as $\hat{\mathbf{H}} \in \mathbb{R}^{L \times d}$.
Subsequently, we introduce dual decoders to reconstruct the raw text and image,  facilitating codebooks to learn the content semantics of the item.

\paratitle{Text Reconstruction.}
Before feeding discrete representations $\hat{\mathbf{H}}$ into the decoder, we first concatenate them with mask tokens: $\mathbf{H}^t = [\hat{\mathbf{H}}, \mathbf{M}^t] $, where $\mathbf{M}^t\in\mathbb{R}^{|T_i|\times d}$ is a mask token matrix used to label the item text to be reconstructed, obtained by repeating a mask embedding $|T_i|$ times.
After that, we introduce a lightweight decoder (\ie one-layer Transformer model) with bidirectional attention to predict the raw item text. 
Formally, the optimization objective for text reconstruction is the following negative log-likelihood loss:
\begin{align}
    \mathcal{L}^t = - \sum_{x \in T_i}\operatorname{log} P(x|\text{Dec}^t(\mathbf{H}^t)),
    \label{eq:dec_text}
\end{align}
where $x$ denotes the token in item text $T_i$, $\text{Dec}^t$ is the text decoder.
The above task is analogous to masked language modeling (MLM)~\cite{bert} with a mask probability of 100\%, which forces the decoder to rely on discrete content representations for text reconstruction. 
Combined with the lightweight and weak decoder, this approach can effectively promote semantic learning of discrete content representations~\cite{retromae}.

\paratitle{Image Reconstruction.}
Similar to text reconstruction, we also concatenate $\hat{\mathbf{H}}$ and the mask matrix $\mathbf{M}^v\in\mathbb{R}^{|V_i|\times d}$ to form vision decoder input $\mathbf{H}^v$.
We then employ a lightweight visual decoder $\text{Dec}^v$ to reconstruct the raw image $V_i$. 
However, representing the raw image as discrete tokens is challenging, as it is more naturally represented by continuous values.
Therefore, we draw upon recent work~\cite{diffloss}, which introduces a small diffusion model and leverages diffusion loss as the reconstruction objective. This approach has been shown to be more advanced than naive MSE loss~\cite{diffloss}.
Thus, the image reconstruction loss is formulated as:
\begin{align}
    \mathcal{L}^v = \text{DiffLoss} (\text{Dec}^v(\mathbf{H}^v), V_i),
    \label{eq:dec_image}
\end{align}
where $\text{DiffLoss}(\cdot, \cdot)$ denotes the diffusion loss, computed based on the decoder's output latent representations, with the raw item image $V_i$ serving as the target.

Combining the above text and image reconstruction losses, we define the following raw content reconstruction loss:
\begin{align}
    \mathcal{L}_{\text{Raw}} = \mathcal{L}^t +\alpha \mathcal{L}^v,
    \label{eq:raw_rec}
\end{align}
where $\alpha$ is a hyper-parameter for the trade-off between text reconstruction and image reconstruction.

\subsubsection{Collaborative Knowledge Integration}
As discussed before, while raw content reconstruction captures universal knowledge from item content information, it overlooks critical collaborative signals that are essential for effective recommendation.
To this end, we adopt the intuitive assumption in recommendation scenarios that co-occurring items are similar.
For each item within the user interaction sequence, we select its one-hop neighbors as positive examples for collaborative knowledge integration. 
By traversing all user interaction sequences, we construct a set of positive instances for item $i$, denoted as $\mathcal{I}^+_i$.
Then, each item is paired with a sampled positive item $i^+ \in \mathcal{I}^+_i$ as input.
For collaborative knowledge integration within the item tokenizer, we propose two tasks: co-occurrence item alignment and reconstruction.

\paratitle{Co-occurring Item Alignment.}
For co-occurring item alignment, we aim to integrate collaborative knowledge into the MLLM encoder by aligning the content representations of positive items.
Formally, as described in Section~\ref{sec:encode}, we obtain the content representations $\mathbf{H}$ and $\mathbf{H}^+$ for the item $i$ and its positive item $i^+$, respectively.
Afterward, we apply contrastive learning to align $\mathbf{H}$ and $\mathbf{H}^+$ through the InfoNCE~\cite{infonce} loss with in-batch negatives.
The co-occurring item alignment loss can be calculated as follows:
\begin{align}
    \mathcal{L}_{\text{Ali}}=- \sum_{l=1}^L \log\frac{\exp(cos(\bm{h}_l, \bm{h}_l^+)/\tau)}{ \sum_{\hat{\bm{h}}_l \in \mathcal{B}_l} \exp(cos(\bm{h}_l, \hat{\bm{h}}_l)/\tau) },
    \label{eq:alignment}
\end{align}
where $\tau$ denotes a temperature hyper-parameter, and $\mathcal{B}_l$ is a batch of content representations for positive items in $l$-th level.
Since the batch data is constructed randomly, the negative instances within a batch are from a mixture of multiple domains, thereby facilitating the fusion of cross-domain knowledge.

\paratitle{Co-occurring Item Reconstruction.}
For co-occurrence reconstruction, our goal is to ensure that discrete content representations of positive pairs capture similar semantics.
As discrete representations are looked up from codebooks and shared across different items, the representations of positive items may also appear in negative examples, which undermines the reliability of contrastive learning.
Therefore, we propose reconstructing the content of the positive item based on discrete content representations $\hat{\mathbf{H}}$ corresponding to the current item.
Formally, the co-occurring item reconstruction loss is calculated as:
\begin{align}
    \mathcal{L}_{\text{Re}} &= \mathcal{L}^{t^+} +\alpha \mathcal{L}^{v^+},
     \label{eq:pos_rec}
\end{align}
where $\mathcal{L}^{t^+}$ denotes the loss for positive item text reconstruction (Eqn.~\eqref{eq:dec_text}), and $\mathcal{L}^{v^+}$ denotes the loss for positive item image reconstruction (Eqn.~\eqref{eq:dec_image}).
This negative-independent approach implicitly encourages the discrete representations of item $i$ to learn consistent semantics of its positive sample, enabling codebooks to effectively encode the semantic similarity between them.

Finally, we integrate the raw content reconstruction loss (Eqn.~\eqref{eq:raw_rec}), the codebook learning loss (Eqn.~\eqref{eq:code_loss}), and the two collaborative knowledge integration losses (Eqn.~\eqref{eq:alignment} and Eqn.~\eqref{eq:pos_rec}) to formulate the overall objective of the universal item tokenizer as follows:
\begin{align}
    \mathcal{L}_{T} = \mathcal{L}_{\text{Raw}} + \lambda \mathcal{L}_{\text{Code}} + \mu \mathcal{L}_{\text{Ali}} + \eta \mathcal{L}_{\text{Re}},
    \label{eq:tokenizer_loss}
\end{align}
where $\lambda$,  $\mu$ and $\eta$ are hyper-parameters for the trade-off between various objectives.

\subsection{Recommender Learning Framework}
\label{sec:training}
After learning a universal item tokenizer, we first pre-train a generative recommender based on multi-domain data, and then fine-tune the pre-trained model for downstream domains.

\subsubsection{Multi-Domain Pre-training}
The multi-domain pre-training for generative recommender consists of two phases:

\paratitle{Multi-Domain Item Tokenization.}
We utilize the pre-trained universal item tokenizer $T$ to uniformly map items across multiple domains to their corresponding identifiers, ensuring that all domains sharing the same code space.
Thus, the item sequence $S$ and the target item $i_{n+1}$ are tokenized as $X=[c^1_1, c^1_2, \dots, c^n_{L-1}, c^n-L]$ and $Y=[c^{n+1}_1, \dots, c^{n+1}_L]$ respectively.
These tokenized sequences from multiple domains are then mixed to form the pre-training data.
Regarding conflict handling, unlike previous works~\cite{tiger,letter} that append a semantically irrelevant extra code, we reassign the last-level code to the second-nearest or a farther neighbor code.
Finally, we mix code sequences from multiple domains as pre-training data for the generative recommender.

\paratitle{Generative Recommender Optimization.}
Generative recommendation reformulates the next-item prediction task as a sequence-to-sequence problem. 
Therefore, we optimize the model by minimizing the negative log-likelihood loss of the target sequence $Y$:
\begin{align}
    \mathcal{L}_R = - \operatorname{log} P(Y|X) =  - \sum_{l=1}^{L}\operatorname{log} P(c^{n+1}_l|X,c^{n+1}_1,\dots,c^{n+1}_{l-1}).
    \label{eq:lmloss}
\end{align}

\subsubsection{Downstream Fine-tuning}
In order to transfer and adapt to new domains, we consider two fine-tuning stages: 

\paratitle{Item Tokenizer Fine-tuning.}
For effective knowledge transfer across multiple domains, a key challenge in generative recommendation is to maintain the correlation and transformation patterns between different codes learned during pre-training.
To this end, we propose fixing the primary parameters (\ie $E^r$ and $E^f$) of the tree-structured codebooks to retain the general knowledge and only fine-tune projection matrices $W^r$ and $W^f$ to incorporate domain-specific adaptations.
Additionally, since the item tokenizer is fine-tuned within a specific domain, we fix the number of item codes $L$ during fine-tuning. 
The fine-tuning task mirrors the pre-training process, and the loss function remains the same as defined in Eqn.~\eqref{eq:tokenizer_loss}. 

\paratitle{Generative Recommender Fine-tuning.}
Given the fine-tuned item tokenizer, domain adaptation of the generative recommender is straightforward.
We first employ the fine-tuned item tokenizer to tokenize the item interaction data from the new domain into code sequences.
Then, the pre-trained generative recommender is fine-tuned using the negative log-likelihood loss defined in Eqn.~\eqref{eq:lmloss}.

%% file: sec3_experiment.tex
\section{Experiments}
\label{sec:experiments}

This section first sets up the experiment and then presents overall performance as well as in-depth analyses of UTGRec.

\subsection{Experiment Setup}

\begin{table}[]
    \centering
    \small
    \caption{Statistics of the preprocessed datasets. Avg.\textit{len} denotes the average length of item sequences.}
    \label{tab:data_statistics}
    \huge
    \resizebox{\linewidth}{!}{
    \begin{tabular}{lrrrrr}
    \toprule
     Dataset    &\#Users   &\#Items   &\#Inter. &Avg.\textit{len} &Sparsity   \\
     \midrule
     Pre-training &999,334 &344,412 &8,609,909 &8.62 &99.997\% \\
     \midrule
     Instrument &57,439  &24,587  &511,836 &8.91 &99.964\%  \\
     Scientific &50,985  &25,848  &412,947 &8.10 &99.969\% \\
     Game &94,762  &25,612  &814,586 &8.60 &99.966\% \\
     Office &223,308  &77,551  &1,577,570 &7.07 &99.991\% \\
     \bottomrule
    \end{tabular}}
\end{table}

\subsubsection{Dataset}

To evaluate the performance of UTGRec, we select nine subsets from Amazon 2023 review dataset~\cite{amazon2023} and divide them into two groups as pre-training datasets and downstream datasets respectively. 
Five domains are used for pre-training: ``Arts Crafts and Sewing'', ``Baby Products'', ``CDs and Vinyl'', ``Cell Phones and Accessories'', and  ``Software''.
Another four subsets as downstream datasets: ``Musical Instruments'', ``Industrial and Scientific'', ``Video Games'', and ``Office Products''.
Following prior studies~\cite{unisrec,tiger}, we apply five-core filtering to all pre-training and downstream datasets, discarding low-activity users and items with fewer than five interactions.
We group the historical item sequences by users and sort them by timestamps, with a maximum sequence length limit of 20 items.
For item text, we concatenate the title, feature, and category fields within item metadata.
For item images, we employ the ``large'' size images provided in the dataset.
The detailed statistics of preprocessed datasets are presented in Table~\ref{tab:data_statistics}.

\subsubsection{Baseline Models}

To enable a comprehensive comparison, we categorize the baseline models into three distinct groups:

\noindent \textbf{(1) Traditional sequential recommenders}:

$\bullet$ {\textbf{GRU4Rec}}~\cite{gru4rec} leverages Gated Recurrent Units (GRUs) to model sequential patterns in user interactions.

$\bullet$ {\textbf{BERT4Rec}}~\cite{bert4rec} employs the bidirectional self-attention mechanism with a masked prediction objective for sequence modeling.

$\bullet$ {\textbf{SASRec}}~\cite{sasrec} utilizes a unidirectional self-attention network to model user behaviors.

$\bullet$ {\textbf{FMLP-Rec}}~\cite{fmlp-rec} introduces an all-MLP model with learnable filters to reduce noise and effectively capture user preferences.

\noindent \textbf{(2) Content-based sequential recommenders}:

$\bullet$ {\textbf{FDSA}}~\cite{fdsa} presents a dual-stream self-attention framework that independently models item-level and feature-level sequences for recommendation.

$\bullet$ {\textbf{S$^3$-Rec}}~\cite{s3rec} enhances sequential recommendation models by leveraging feature-item correlations as self-supervised signals.

$\bullet$ {\textbf{UniSRec}}~\cite{unisrec} leverages the associated description text of items to learn universal representations across different domains. 

$\bullet$ {\textbf{VQ-Rec}}~\cite{vqrec} proposes representing item text as discrete codes and subsequently learning transferable code embeddings for universal sequence modeling.

$\bullet$ {\textbf{MISSRec}}~\cite{missrec} learns multimodal interest-aware sequence representation for transferable recommendation.

\noindent \textbf{(3) Generative recommenders}:

$\bullet$ {\textbf{TIGER}}~\cite{tiger} employs RQ-VAE to map items into semantic IDs, which serve as item identifiers, and adopts the generative retrieval paradigm for sequential recommendation.

$\bullet$ {\textbf{LETTER}}~\cite{letter} enhances TIGER by incorporating collaborative and diversity regularization into RQ-VAE.

$\bullet$ {\textbf{TIGER}$_M$}~\cite{tiger} utilizes multimodal embeddings as item semantic embeddings for training the RQ-VAE. For the multimodal encoder, we tried CLIP~\cite{clip} and Qwen2-VL-2B~\cite{qwen2-vl}. 
The results reported are the best of these two encoders.

\subsubsection{Evaluation Settings}

We evaluate model performance using two widely adopted metrics: top-$K$ Recall and Normalized Discounted Cumulative Gain (NDCG), with $K$ set to 5 and 10.
Following previous studies~\cite{sasrec, s3rec, tiger}, we employ the \emph{leave-one-out} strategy for dataset splitting. 
For each user interaction sequence, the latest item is designated as the test data, the second most recent item as the validation data, and all remaining items as the training data. 
To ensure a rigorous comparison, we conduct the full-ranking evaluation over the entire item set.
Additionally, the beam size is set to 50 for all generative recommendation models.

\subsubsection{Implementation Details}
\label{sec:implement}

We utilize Qwen2-VL-2B~\cite{qwen2-vl} as the backbone for our universal item tokenizer and perform low-rank fine-tuning using LoRA~\cite{lora}. 
The dual decoders are implemented as two one-layer Transformer models.
The root and leaf codebook sizes are set to 256 and 512, respectively, maintaining the total number of learnable codebook vectors consistent with baseline methods (\eg TIGER). The item identifier length $L$ is set to 3.
For tokenizer optimization, we use the AdamW optimizer with initial learning rates of 3e-4 for pre-training and 1e-4 for fine-tuning.
A cosine scheduler is employed to dynamically adjust the learning rates.
The batch size per GPU is set to 16, with 16 GPUs used for pre-training over 3 epochs, followed by fine-tuning on downstream datasets for 20 epochs. 
The loss coefficient $\lambda$ for codebook learning is set to 200.
The hyper-parameters $\alpha$, $\mu$, and $\eta$ are tuned in \{0.3, 1, 3, 5, 10\}, \{0.001, 0.003, 0.01, 0.03\}, and \{0.01, 0.03, 0.1, 0.3\}, respectively.
Furthermore, all hyper-parameters are kept consistent across both pre-training and fine-tuning phases.
For the generative recommender, we adopt the settings in TIGER~\cite{tiger}, which employs T5~\cite{t5} as the backbone, and we tune the number of encoder and decoder layers within \{1, 2, 3, 4, 5, 6\}.
We also leverage the AdamW optimizer and cosine scheduler to optimize the recommender. 
The initial learning rates for pre-training and fine-tuning are set to 0.005 and 0.003, respectively.
The batch size per GPU is 256 and 4 GPUs are used for recommender training. 
The recommender is pre-trained across multiple domains for 50 epochs, while fine-tuning is conducted using an early stopping strategy based on validation performance to ensure convergence and prevent overfitting.

\begin{table*}
\centering
\caption{The overall performance comparisons between different baseline methods and UTGRec. The best and second-best results are highlighted in bold and underlined font, respectively. ``*'' denotes that the improvements are statistically significant with $p<0.01$ in a paired t-test setting.}
\label{tab:main_res}
\huge
\resizebox{\textwidth}{!}{
\renewcommand\arraystretch{1.1}
\begin{tabular}{l|l|cccc|ccccc|cccc}
\hline
Dataset & Metric    & GRU4Rec & BERT4Rec & SASRec & FMLP-Rec & FDSA   & S$^3$-Rec  & UniSRec & VQ-Rec & MISSRec & TIGER  & LETTER & TIGER$_M$ & UTGRec \\
\hline
\multirow{4}{*}{Instrument} & Recall@5  & 0.0324  & 0.0307   & 0.0333 & 0.0339   & 0.0347 & 0.0317 & 0.0356  & 0.0361 & \underline{0.0384}  & 0.0370 & 0.0372 & 0.0365  &  \textbf{0.0398*}      \\
                            & NDCG@5    & 0.0209  & 0.0195   & 0.0213 & 0.0218   & 0.0230 & 0.0199 & 0.0228  & 0.0231 & \underline{0.0250}  & 0.0244 & 0.0246 & 0.0239  &  \textbf{0.0263*}      \\
                            & Recall@10 & 0.0501  & 0.0485   & 0.0523 & 0.0536   & 0.0545 & 0.0496 & 0.0560  & 0.0578 & \underline{0.0590}  & 0.0564 & 0.0580 & 0.0551  &  \textbf{0.0616*}      \\
                            & NDCG@10   & 0.0266  & 0.0252   & 0.0274 & 0.0282   & 0.0293 & 0.0257 & 0.0299  & 0.0301 & 0.0307  & 0.0306 & \underline{0.0313} & 0.0298  &  \textbf{0.0334*}      \\
\hline
\multirow{4}{*}{Scientific} & Recall@5  & 0.0202  & 0.0186   & 0.0259 & 0.0269   & 0.0262 & 0.0263 & 0.0276  & 0.0285 & \underline{0.0291}  & 0.0264 & 0.0279 & 0.0280  &  \textbf{0.0308*}      \\
                            & NDCG@5    & 0.0129  & 0.0119   & 0.0150 & 0.0155   & 0.0169 & 0.0171 & 0.0174  & 0.0188 & \underline{0.0190}  & 0.0175 & 0.0182 & 0.0182  &  \textbf{0.0204*}      \\
                            & Recall@10 & 0.0338  & 0.0296   & 0.0412 & 0.0422   & 0.0421 & 0.0418 & 0.0437  & 0.0444 & \underline{0.0452}  & 0.0422 & 0.0435 & 0.0437  &  \textbf{0.0481*}      \\
                            & NDCG@10   & 0.0173  & 0.0155   & 0.0199 & 0.0204   & 0.0213 & 0.0219 & 0.0226  & 0.0229 & \underline{0.0237}  & 0.0226 & 0.0232 & 0.0233  &  \textbf{0.0255*}      \\
\hline
\multirow{4}{*}{Game}       & Recall@5  & 0.0499  & 0.0460   & 0.0535 & 0.0528   & 0.0544 & 0.0485 & 0.0546  & 0.0560 & 0.0570  & 0.0559 & 0.0563 & \underline{0.0571}  & \textbf{0.0592*}       \\
                            & NDCG@5    & 0.0320  & 0.0298   & 0.0331 & 0.0338   & 0.0361 & 0.0315 & 0.0353  & 0.0333 & 0.0337  & 0.0366 & 0.0372 & \underline{0.0375}  & \textbf{0.0390*}       \\
                            & Recall@10 & 0.0799  & 0.0735   & 0.0847 & 0.0857   & 0.0852 & 0.0769 & 0.0864  & 0.0879 & 0.0862  & 0.0868 & \underline{0.0877} & 0.0873  &   \textbf{0.0909*}     \\
                            & NDCG@10   & 0.0416  & 0.0386   & 0.0438 & 0.0444   & 0.0448 & 0.0406 & 0.0453  & 0.0443 & 0.0426  & 0.0467 & \underline{0.0473} & 0.0472  &    \textbf{0.0491*}    \\
\hline
\multirow{4}{*}{Office}     & Recall@5  & 0.0204  & 0.0182   & 0.0255 & 0.0259   & 0.0263 & 0.0247 & 0.0246  & 0.0272 & 0.0264  & 0.0288 & \underline{0.0290} & 0.0280  & \textbf{0.0320*}       \\
                            & NDCG@5    & 0.0135  & 0.0119   & 0.0149 & 0.0156   & 0.0177 & 0.0139 & 0.0160  & 0.0178 & 0.0166  & 0.0199 & \underline{0.0201} & 0.0195  &    \textbf{0.0224*}     \\
                            & Recall@10 & 0.0307  & 0.0277   & 0.0375 & 0.0392   & 0.0380 & 0.0371 & 0.0372  & 0.0401 & 0.0374  & 0.0417 & \underline{0.0421} & 0.0400  &   \textbf{0.0462*}     \\
                            & NDCG@10   & 0.0168  & 0.0150   & 0.0187 & 0.0199   & 0.0215 & 0.0179 & 0.0201  & 0.0219 & 0.0200  & 0.0241 & \underline{0.0243} & 0.0234  &  \textbf{0.0269*}     \\
\hline
\end{tabular}}
\end{table*}

\subsection{Overall Performance}

We compare UTGRec with various baseline models on four public recommendation benchmarks. The overall results are shown in Table~\ref{tab:main_res}. From these results, we have the following observations:

Compared to traditional recommendation models (\ie GRU4Rec, BERT4Rec, SASRec, FMLP-Rec), content-based models (\ie FDSA, S$^3$-Rec) achieve superior results and pre-training across multiple domains leads to further improvements (\ie UniSRec, VQ-Rec, MISSRec).
This phenomenon shows that incorporating item content information (\eg item texts and images) and learning universal item representations can significantly improve recommendation performance.
Generative recommenders (\ie TIGER, LETTER, TIGER$_M$) benefit from the generative paradigm and the prior semantics within item identifiers, outperforming traditional recommendation models.
Among these, LETTER surpasses TIGER due to its application of collaborative and diversity regularization in the item tokenizer.
TIGER$_M$ improves upon TIGER on Scientific and Game datasets but underperforms on Instrument and Office datasets, which indicates that a naive integration of multimodal features may not always yield beneficial outcomes.
Furthermore, limited by the domain-specific tokenizers and recommenders, generative baseline models do not consistently perform better than transferable sequential recommenders (\eg MISSRec). 
This further highlights the importance of developing a transferable generative recommender.


Finally, our proposed UTGRec maintains the best performance in all cases, exhibiting substantial improvements over traditional, content-based, and generative baseline models.
Different from previous generative recommenders, we propose a universal item tokenizer that leverages multimodal content for item semantic modeling.
Based on this, we develop a transferable generative recommender, which effectively enhances model performance by integrating the strengths of the generative paradigm and cross-domain knowledge transfer.

\begin{table}[]
\centering
\caption{Ablation study of our approach.}
\label{tab:ablation}
\resizebox{1.0\linewidth}{!}{
\begin{tabular}{lcccc}
\toprule
\multirow{2}{*}{Methods} & \multicolumn{2}{c}{Instrument} & \multicolumn{2}{c}{Scientific} \\
\cmidrule(l){2-3} \cmidrule(l){4-5}
                        & Recall@10       & NDCG@10      & Recall@10       & NDCG@10      \\
\hline
(0)   UTGRec     & \textbf{0.0616}	& \textbf{0.0334}	& \textbf{0.0481}	& \textbf{0.0255}  \\
(1)  \  w/o TreeCode  & 0.0591	& 0.0319	& 0.0457	& 0.0239         \\
(2)  \  w/o $\mathcal{L}_{\text{Ali}}$  & 0.0599	& 0.0321	& 0.0464	& 0.0243         \\
(3) \   w/o $\mathcal{L}_{\text{Re}}$   & 0.0607	& 0.0326	& 0.0468	& 0.0246       \\
(4) \  w/o FT(T)    & 0.0551	& 0.0302	& 0.0415	& 0.0218 \\
(5) \ Full FT      & 0.0559	& 0.0304	& 0.0423	& 0.0225   \\
(6)  \ w/o PT(R)   & 0.0583	& 0.0315	& 0.0451	& 0.0236 \\
(7) \ w/o PT      & 0.0575	& 0.0310	& 0.0444	& 0.0232   \\
\bottomrule
\end{tabular}}
\end{table}

\subsection{Ablation Study}
To investigate how the proposed techniques impact model performance, we conduct an ablation study on Instrument and Scientific datasets. Specifically, we consider the following six variants of UTGRec:
(1) \underline{w/o TreeCode} without the tree-structured codebooks and instead applying the multi-level codebooks like RQ-VAE (Section~\ref{sec:tree_code}).
(2) \underline{w/o $\mathcal{L}_{\text{Ali}}$} without the co-occurring item alignment loss (Eqn.~\eqref{eq:alignment}).
(3) \underline{w/o $\mathcal{L}_{\text{Re}}$} without the co-occurring item  reconstruction loss (Eqn.~\eqref{eq:pos_rec}).
(4) \underline{w/o FT(T)} without tokenizer fine-tuning when transferring to downstream domains, instead directly using the pre-trained universal item tokenizer to tokenize items in the new domain.
(5) \underline{Full FT} does not fix the codebook matrices (\ie $E^r$ and $E^f$) of the tree-structured codebooks during fine-tuning.
(6) \underline{w/o PT(R)} without generative recommender pre-training. On downstream datasets, only the pre-trained item tokenizer is fine-tuned, while the generative recommender is trained from scratch. 
(7) \underline{w/o PT} without both tokenizer and recommender pre-training, but learns both components within a domain-specific setting.

The experimental results for our UTGRec and its multiple variants are presented in Table~\ref{tab:ablation}.
As observed, removing any of the aforementioned techniques leads to a decline in overall performance.
The tree-structured codebooks outperform the straightforward multi-level codebooks (\ie variant (1)).
The absence of the co-occurring item alignment and reconstruction (\ie variants (2) and (3)) results in a lack of collaborative knowledge within the universal item tokenizer, causing performance degradation.
Directly applying the pre-trained item tokenizer to downstream datasets (\ie variant (4)) fails to achieve the expected results, which is likely due to the entanglement of item codes in the new domain.
Furthermore, fine-tuning all codebook parameters (\ie variant (5)) may disrupt the associations between codes, resulting in the loss of general knowledge acquired during pre-training.

\subsection{Further Analysis}

\begin{figure}[]
\centering
\includegraphics[width=0.99\linewidth]{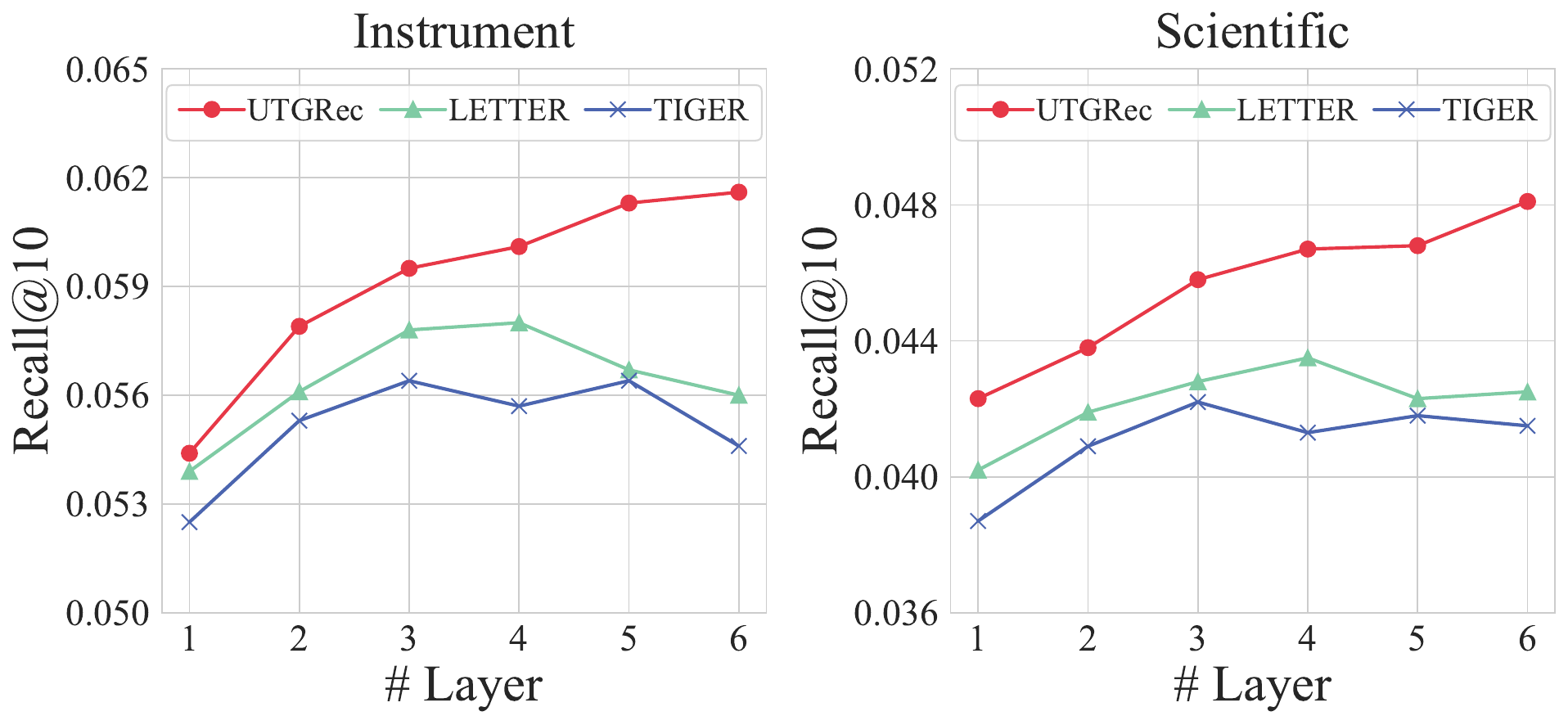}
\caption{Performance comparison \wrt model scale. }
\label{fig:model_scale}
\vspace{-5pt}
\end{figure}

\begin{figure}[]
\centering
\includegraphics[width=0.99\linewidth]{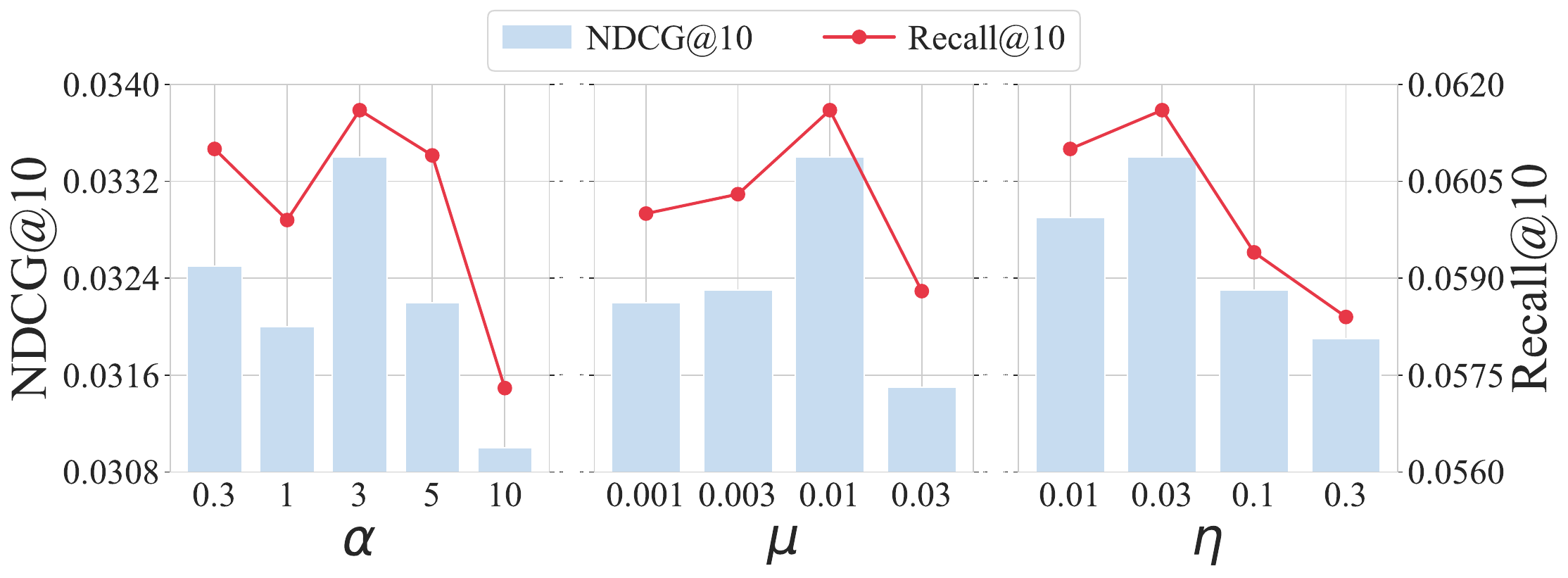}
\caption{Hyper-parameter tuning on instrument dataset. }
\label{fig:hyper}
\vspace{-5pt}
\end{figure}

\subsubsection{Recommender Scalability Analysis}

The generative recommender for a specific domain is often limited by data sparsity in recommendation scenarios, making it difficult to improve the performance through model scaling, as seen with LLMs~\cite{llm_survey, scaling_instruct, scaling_law}. 
In contrast, our proposed universal item tokenizer allows us to leverage more massive and diverse data from multiple domains, thereby facilitating the effective scaling of generative recommenders.
In this part, we explore the scalability of UTGRec by gradually increasing the number of encoder and decoder layers of the generative recommender to six layers.
From the results in Figure~\ref{fig:model_scale}, we can observe that the performance of baseline models (\ie TIGER, LETTER) shows a positive correlation with model scale only in a few layers.  
With a slight increase in model size, performance begins to degrade due to overfitting.
Conversely, the performance of UTGRec generally improves as the model scales, which demonstrates that universal item tokenization and recommender pre-training are beneficial to the scalability of generative models.

\subsubsection{Hyper-Parameter Analysis}

We continue to investigate the impact of hyper-parameters in our approach.
For the loss coefficient of image reconstruction, we tune $\alpha$ within the range \{0.3, 1, 3, 5, 10\}.
As shown in Figure ~\ref{fig:hyper}, the model attains the best results by achieving a trade-off between text and image semantic learning when $\alpha$ is set to 3. 
For the loss coefficient of co-occurring item alignment, we tune $\mu$ within the ranges \{0.001, 0.003, 0.01, 0.03\}.
The results indicate that an inappropriate $\mu$ impairs the learning of item content representations, leading to suboptimal performance. 
The optimal value of $\mu$ is found to be 0.01.
For the loss coefficient of co-occurring item reconstruction, we explore values of $\eta$ from the set \{0.01, 0.03, 0.1, 0.3\}.
UTGRec exhibits subpar performance when $\eta$ is overly large and achieves optimal performance when $\eta$ is 0.03.

\begin{figure}[]
\centering
\includegraphics[width=0.99\linewidth]{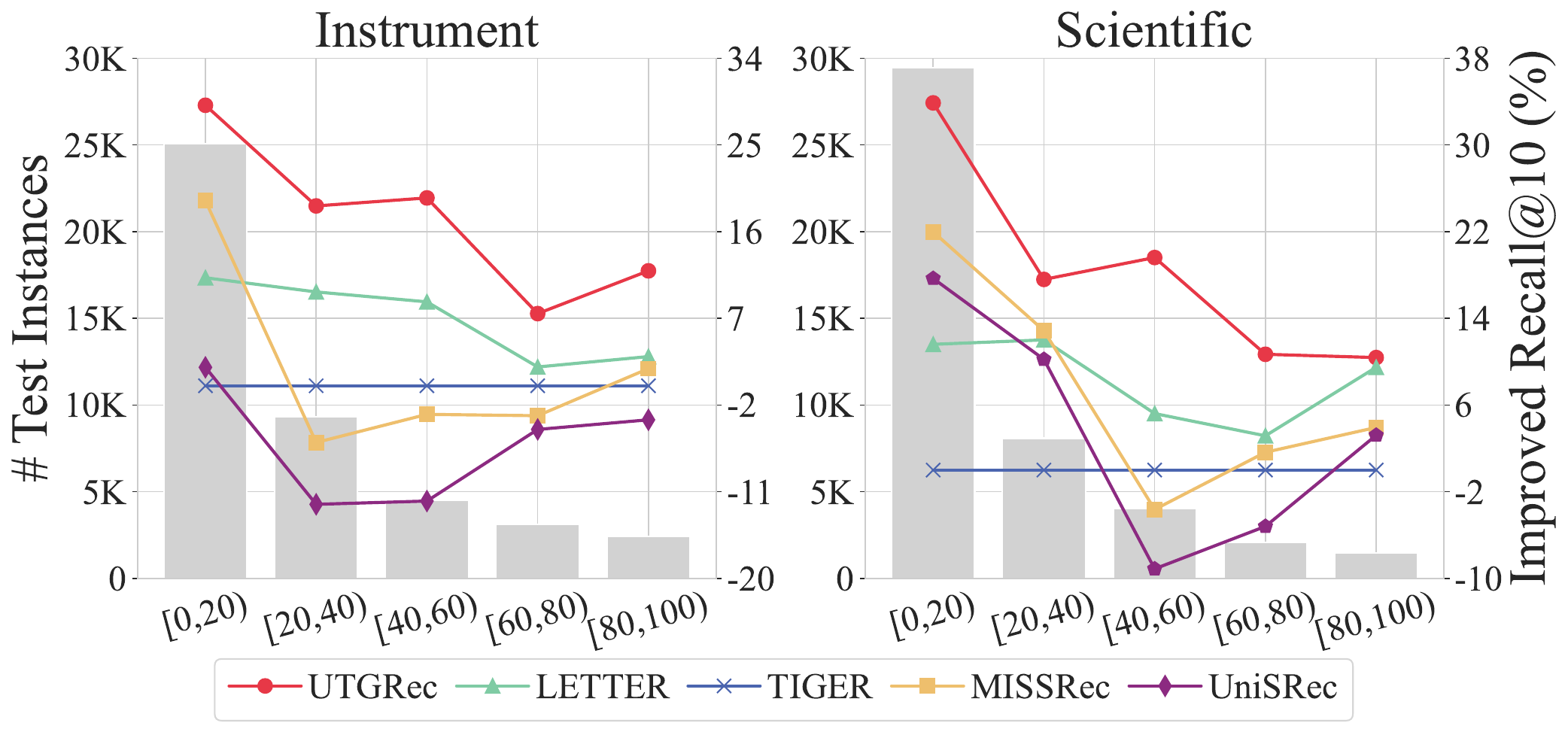}
\caption{Performance comparison \wrt long-tail items. }
\label{fig:long_tail}
\end{figure}

\subsubsection{Performance Comparison \wrt Long-tail Items}

To validate the universality and generality on long-tail items, we split the test data into different groups according to the popularity of target items.
Then, we compare the performance of various models on these data groups and present the improvements over TIGER in Figure~\ref{fig:long_tail}.
Transferable sequential recommenders (\ie UniSRec, MISSRec) outperform TIGER in the cold-start groups (\eg group [0, 20)), but their performance is weaker compared to generative models in popular groups (\eg group [40, 60)).
In contrast, UTGRec consistently surpasses the baseline models across all groups, especially in less popular groups.
This shows significant advantages of our transferable approach on long-tail items. 



%% file: sec4_relatedwork.tex
\section{Related Work}
\label{sec:related}
In this section, we review related work from three aspects, sequential recommendation, generative recommendation and transfer learning for recommendation.

\paratitle{Sequential Recommendation.}
Sequential recommendation aims to predict the next potential item by capturing user's personalized preferences in his/her sequential behaviors.
Early studies~\cite{fpmc,fossil} follow the Markov chain assumption and learn the transformation matrix between the items.
With the rapid development of deep learning, 
recent works typically leverage various deep neural networks to capture the sequential patterns of item ID sequence, including convolutional neural network (CNN)~\cite{caser}, recurrent neural network (RNN)~\cite{gru4rec,DBLP:conf/recsys/TanXL16}, graph neural network (GNN)~\cite{DBLP:conf/aaai/WuT0WXT19,DBLP:conf/ijcai/XuZLSXZFZ19}, multilayer perceptron (MLP)~\cite{fmlp-rec} and Transformer~\cite{sasrec,bert4rec}.
Furthermore, several studies introduce item attributes~\cite{fdsa,s3rec} or self-supervised signals~\cite{s3rec,cl4srec,duorec} to enhance the sequence modeling.
However, these methods are mainly developed based on item IDs, which results in item embeddings and model parameters being domain-specific.

\paratitle{Generative Recommendation.}
Generative recommendation~\cite{tiger} is a promising paradigm that reformulates the next-item prediction task as a sequence-to-sequence problem.
In this paradigm, each item is tokenized into an identifier composed of multiple codes, and then a generative model is employed to autoregressively generate the target item identifier.
Existing approaches for item tokenization can be roughly categorized into three groups: heuristic methods, cluster-based methods, and codebook-based methods.
Heuristic methods primarily rely on manually defined rules, such as time order~\cite{p5} and item category~\cite{howtoindex}, to generate item identifiers. 
Cluster-based methods group items based on their embeddings~\cite{seater,eager} or the co-occurrence matrix~\cite{gptrec,howtoindex} to assign item identifiers.
Codebook-based methods~\cite{tiger,ding2024inductive,DBLP:journals/corr/abs-2403-18480, tokenrec,sc,letter,etegrec} adopt learnable codebooks to quantize item semantic embeddings, thereby constructing fixed-length and semantically rich item identifiers.
However, these methods mainly focus on learning domain-specific item tokenizers, which limits the ability of generative recommenders to transfer across domains. 

\paratitle{Transfer Learning for Recommendation.}
Transferring knowledge from other domains~\cite{DBLP:conf/kdd/Xie0WLZL22,DBLP:conf/cikm/ZhuC0LZ19,DBLP:conf/ijcai/ZhuW00L021} is a widely adopted strategy to address data sparsity and cold-start issues in recommender systems. 
Traditional transfer learning methods primarily rely on shared information between the source and target domains, such as overlapping users~\cite{DBLP:conf/cikm/HuZY18,DBLP:conf/emnlp/WuWQLH020,DBLP:conf/sigir/YuanZKJKL21}, items~\cite{DBLP:conf/kdd/SinghG08,DBLP:conf/cikm/ZhuC0LZ19}, or attributes~\cite{DBLP:conf/kdd/TangWSS12}.
Recently, several studies~\cite{unisrec,vqrec,missrec} employ pre-trained language~\cite{bert,sentence-t5} or multimodal models~\cite{clip} to learn universal item representation and enable transferable sequential recommendation without overlapping users or items.
However, these methods are still rooted in the traditional sequential recommendation paradigm, which is implemented by measuring the similarity between the sequence representation and candidate item representations.
In contrast, we introduce universal item tokenization to achieve more effective transferable recommendations within the generative recommendation paradigm.

%% file: sec5_conclusion.tex
\section{Conclusion}
\label{sec:conclusion}

In this paper, we proposed UTGRec, a novel framework which achieves transferable generative recommendation through universal item tokenization.
Unlike previous methods that learn domain-specific tokenizers, we introduced a universal item tokenizer that employs a MLLM with tree-structured codebooks for item tokenization.
To effectively train the universal item tokenizer, we presented an item content reconstruction approach with collaborative integration.
The reconstruction objective uses the raw item content as the target, enabling the learning of more essential  characteristics. 
Simultaneously, we integrated collaborative knowledge through co-occurring item alignment and reconstruction.
Given the universal item tokenizer, we developed a learning framework to pre-train a transferable generative recommender.
Extensive experiments on four public datasets showed that UTGRec consistently outperforms both traditional and generative recommendation baselines.
For future work, we will incorporate more domains and interaction data.
Additionally, we will also investigate the scaling effect of the item tokenizer and the generative recommender.